# The Franzese-Stanley Coarse Grained Model for Hydration Water

Luis Enrique Coronas[1], Oriol Vilanova[1], Valentino Bianco[2],
Francisco de los Santos[3] and Giancarlo Franzese[1,*]

[1]Departament de Física de la Matèria Condensada & Institut de Nanociència i Nanotecnologia (IN2UB), Universitat de Barcelona, Carrer Martí i Franquès 1, 08028 Barcelona (Spain)

[2]Departamento de Química Física, Universidad Complutense de Madrid, Plaza de las Ciencias, Ciudad Universitaria, 28040 Madrid (Spain)

[3]Departamento de Electromagnetismo y Física de la Materia, Universidad de Granada, Fuentenueva s/n, 18071 Granada (Spain)

* gfranzese@ub.edu

## Abstract

Water modeling is a challenging problem. Its anomalies are difficult to reproduce, promoting the proliferation of a large number of computational models, among which researchers select the most appropriate for the property they study. In this chapter, we introduce a coarse-grained model introduced by Franzese and Stanley (FS) that accounts for the many-body interactions of water. We review mean-field calculations and Monte Carlo simulations on water monolayers for a wide range of pressures and temperatures, including extreme conditions. The results show the presence of two dynamic crossovers and explain the origin of diffusion anomalies. Moreover, the model shows that all the different scenarios, proposed in the last decades as alternative explanations of the experimental anomalies of water, can be related by the fine-tuning of the many-body (cooperative) interaction. Once this parameter is set from the experiments, the FS model predicts a phase transition between two liquids with different densities and energies in the supercooled water region, ending in a liquid-liquid critical point. From this critical point stems a liquid-liquid Widom line, i.e., the locus of maxima of the water correlation length, which in the FS model can be directly calculated. The results are consistent with the extrapolations from experiments. Furthermore, they agree with those from atomistic models but make predictions over a much wider thermodynamic region, allowing for a better interpretation of the available experimental data. All these findings provide a coherent picture of the properties of water and confirm the validity of the FS model that has proved to be useful for large-scale simulations of biological systems.

## 1. Introduction

Water has more than 60 thermodynamic, dynamic and structural anomalies (Chaplin 2006) whose origin is largely debated (Amann-Winkel et al. 2016, Franzese and Stanley 2010, Gallo et al. 2016, Handle et al. 2017). For example, in contrast with normal fluids, water has the property of *polyamorphism*, i.e., it has at least three amorphous solids (Amann-Winkel et al. 2016, Mishima 1994, Mishima et al. 1984, 1985), whose formation depends on the preparation route (Handle et al. 2017), and a large number of (crystal) ice polymorphs: 17 have been confirmed experimentally and other are predicted computationally (Salzmann 2019). It can be supercooled in its liquid state almost 50 degrees below its melting temperature (Kim et al. 2017). Ice has a lower density than liquid water at ambient pressure and its density decreases below 4$^o$C. Experiments (Angell et al. 1973, 1982, Speedy and Angell 1976) have shown that water isobaric specific heat $C_P$ and isothermal compressibility $K_T$ have a non-monotonic behavior with minima at approximately 35$^o$C and 46$^o$C, respectively, at ambient pressure, while isobaric thermal expansivity $\alpha_P$ turns negative at approximately 4$^o$C. All the anomalies of water become more relevant in the supercooled region, where the fluctuations increase upon cooling (Speedy and Angell 1976, Stanley et al. 1981, Gallo et al. 2016) instead of decreasing as in normal liquids.

Several thermodynamic scenarios have been proposed to explain the origin of the anomalies and polyamorphism. The *singularity free* (SF) scenario hypothesizes that the uncommon volume-entropy and volume-energy anticorrelations of water, due to the water hydrogen bonds (HBs) properties, are responsible for the increase of fluctuations in the supercooled region with no singular behavior (Sastry et al. 1996, Stanley and Teixeira 1980). Three other scenarios, the *stability limit* (SL) conjecture (Speedy 1982), the *liquid-liquid critical point* (LLCP) scenario (Poole et al. 1992), and the *critical point free* (CPF) hypothesis (Angell 2008, Poole et al. 1994), instead postulate a singular behavior that enhances the fluctuations at low temperature: A reentrant spinodal for the first, a critical point for the second, and a first-order phase transition for the third scenario. Stokely et al. demonstrated that all the scenarios belong to the same theoretical framework and that it is possible to go from one to another by tuning a single parameter related to the water's cooperativity (Stokely et al. 2010a), as we will discuss in the following. Nevertheless, which of these scenario holds for water is still a matter of debate, because, so far, no definitive experimental evidence has been found, although many recent experiments are contributing to enrich our insight (Caupin et al. 2018, Kim et al. 2017, 2018, Woutersen et al. 2018). One of the issues is that water freezes before experimental measurements are made in the region, conventionally called 'no-man's land' (Handle et al. 2017, Mishima and Stanley 1998, Stanley et al. 2005), where the different scenarios predict different behaviors. Several strategies have been explored to overcome the inevitable crystallization of supercooled water, including strong confinement (Liu et al. 2007, Mallamace et al. 2007a,b) and anti-freezing solutions (Murata and Tanaka 2012, 2013, Woutersen et al. 2018) with results possibly related to the bulk case (Leoni and Franzese 2016a, Mancinelli et al. 2009, Soper 2012, Wang et al. 2016).

In our research, we use all-atom simulations to study systems where water is at an interface, forming a monolayer or a few confined layers. Experiments (Paul 2012, Zhang et al. 2011) and simulations on confined water (Camisasca et al. 2017, Han et al. 2010, Zangi and Mark 2003) show controversial phenomena, as water under confinement can have properties significantly different from those found in bulk water. For instance, experiments on water confined between flat crystals of graphite and hexagonal boron nitride find a decay of two orders of magnitude in the dielectric constant of water as a function of the sample thickness (down to 1 nanometer) compared to the bulk case (Fumagalli et al. 2018). The mobility of water is also strongly affected by confinement, as different regimes can be observed depending on the structure, composition, and geometry of the confining surfaces, going from subdiffusive to superdiffusive (Gallo et al. 2010, 2000, Leoni and Franzese 2016b). Moreover, confinement also modifies the phase diagram, as crystallization can be avoided, at least partially, at temperatures down to 160 K for water confined in silica MCM-41 nanopores (Faraone et al. 2004, Leoni and Franzese 2016a, Stefanutti et al. 2019). Confined water is also of great importance in biological systems. For example, hydration water plays an important role in protein denaturation (Bianco and Franzese 2015, Bianco et al. 2017a,b, Piana and Shaw 2018) and protein aggregation (Bianco et al. 2019a,b). Water between cell membranes is responsible for self-assembly of phospholipids into bilayers (Nagle and Tristram-Nagle 2000, Zhao et al. 2008) and stabilizes the membrane structure (Calero and Franzese 2019, Martelli et al. 2017, Samatas et al. 2019). Therefore, it is clear that a deeper understanding of confined water properties is essential to develop new applications in bionanotechnology.

We study hydrated systems over time-scales going from 10 fs to 1 $\mu$s and length-scales from 1 Å to 100 nm, adopting atomistic water models such as SCP/E, ST2, TIP3P, TIP4P/2005 and TIP5P with numbers of water molecules up to 7040 (Calero and Franzese 2019, Calero et al. 2016, Kesselring et al. 2012, 2013, Kumar et al. 2006a, Martelli et al. 2017, Martí et al. 2017, Samatas et al. 2019). Nevertheless, these models have extremely large equilibration times at very low temperature or extreme pressure (González et al. 2016, Vega et al. 2009, Yagasaki et al.2014) and have free-energy minima that differ from those of more realistic polarizable models (Hernández-Rojas et al. 2010, James et al. 2005). We, therefore, use the all-atoms results to define coarse-grained models for water monolayers (Franzese et al. 2003, Franzese and Stanley 2002a,b, 2007, 2010, Franzese et al. 2008, 2000, Kumar et al. 2008a,b,c, Mazza et al. 2009, Stokely et al. 2010a,b) hydrating proteins or nanoparticles (Bianco and Franzese 2014, 2015, Bianco et al. 2017a, 2012b, 2013a, 2017b, 2013b, Franzese and Bianco 2013, Franzese et al. 2011, 2010, Mazza et al. 2011, 2012, Strekalova et al. 2011, 2012b). With the coarse-grained models, we can study the large-scale dynamics of hydration water at extreme conditions or the protein folding, spanning scales from 1 nm to 1 $\mu$m in space and from 10 ns to 0.1 s in time.

This chapter reviews our contribution to the ongoing debate on the origin of the water properties adopting the Franzese-Stanley coarse-grained model for a water monolayer (Franzese and Stanley 2002a,b). The model offers a rationale for the thermodynamic and dynamic anomalies of water, relating the cooperativity of the hydrogen bond network to the occurrence of a possible liquid-liquid phase transition (LLPT) in the supercooled water region.

## 2. Computational Models: All Atoms Simulations

Given the experimental difficulties to avoid crystallization of water at extreme low temperatures, it is interesting to resort to computational models to understand if the different theoretical scenarios are, at least, thermodynamically consistent. However, modeling water is a difficult problem (Barnes et al. 1979, Finney 2001), in particular, because it is not settled how to include the quantum many-body, or cooperative, effects in water interactions. In literature, there are more than a hundred water models (Ouyang and Bettens 2015), from those parametrized based on experimental data to those fitting ab-initio calculations, and each model is coarse-grained at a different level, from atomistic non-polarizable and rigid, to polarizable or flexible, from spatially resolved to spatially coarse-grained. For a number of these models, it is possible to explore the *no-man's land*, although it could require extremely large computational times (Kesselring et al. 2012) and elaborated analysis (Chandler 2016, Limmer and Chandler 2013, Palmer et al. 2018, 2014, 2016).

Specifically, for those models belonging to the family of ST2, TIP4P and TIP5P potentials (Gallo et al. 2016), the LLCP hypothesis is, among the different scenarios, the one that better adjusts to the low-temperature phase diagram. In particular, we have shown that the ST2 model has a LLCP between two metastable liquid phases, low-density liquid (LDL) and high-density liquid (HDL), that belongs to the 3D Ising universality class (Kesselring et al. 2012, Lascaris et al. 2013). To this goal, we have performed all-atom simulations and a detailed finite-size scaling analysis based on an appropriate order parameter, defined as a combination of energy and density (Wilding and Binder 1996). Furthermore, we have shown that structural parameters that can quantify the amount of diamond structure in the first shell and the amount of hcp structure in the second shell (Kesselring et al. 2013) discriminate better the different local structures of LDL and HDL.

Similar conclusions have been reached for different models, for example recently in Ref. (Shi et al. 2018), with results that are in principle model dependent. Therefore, understanding which feature of these models regulates the occurrence of anomalies of water and its peculiar properties is a task that requires a detailed analysis for each of them (Shi et al. 2018).

In the following, we will describe our approach to define a model of water that
· is suitable for large-scale, long-time simulations as needed in biologically relevant problems;
· is manageable for theoretical calculations, essential for understanding general properties;

· includes many-body interactions, necessary for a proper water model;
· equilibrates at extreme conditions.

### 3. The spatially coarse-grained Franzese-Stanley Water Model

In their seminal paper about computer simulations of liquid water, Barnes, Finney Nicholas, and Quinn stated that the use of pair-additive interactions for water is a "serious oversimplification" that "is necessary to abandon" because it does not account for many-body forces, to which "solution and interfacial properties of aqueous systems are particularly sensitive" (Barnes et al. 1979). For example, they reviewed quantum calculations for small water clusters, showing that HB energy in trimers and tetramers is 20-30% stronger than in dimers. More recently, James et al. (2005) and Hernández-Rojas et al. (2010) studied the configurations that minimize the energy of water clusters made of up to 21 molecules adopting non-polarizable and polarizable models. The comparison shows structural differences for clusters with more than five molecules, in particular with six or more than ten. This observation emphasizes that many-body effects in water are especially important when there are at least five molecules.

To account for these many-body effects, in 2002 Franzese and Stanley (FS), proposed a coarse-grained Hamiltonian model[1] that is analytically tractable (Franzese and Stanley 2002a,b, 2007) and is suitable for Monte Carlo (MC) calculations at constant pressure $P$ and temperature $T$ (de los Santos and Franzese 2009, Franzese and de los Santos 2009, Franzese et al. 2003, 2008, Kumar et al. 2008a,b,c). Thanks to a percolation mapping (Bianco and Franzese 2019) and a very efficient *cluster* MC algorithm (Franzese et al. 2010, Mazza et al. 2009, Stokely et al. 2010a), the model can be equilibrated at extreme $T < 125$ K and $P$ ranging from negative to more than 10 GPa (Bianco et al. 2012a, 2013a, de los Santos and Franzese 2011, 2012, Mazza et al. 2012, Strekalova et al. 2011, 2012a,b). Furthermore, the coarse-graining allows the simulation of large systems, with more than 160.000 molecules (Bianco and Franzese 2014), a size that is challenging for atomistic models.

On the other hand, by adopting *local* MC dynamics (Kumar et al. 2008c) and a rescaling of time units based on experimental data, it is possible to simulate the model for times up to 100 s (Mazza et al. 2011). This timescale is out of reach for any atomistic simulation and is accessible only for coarse-grained models.

Furthermore, the use of *diffusive* MC dynamics allows evaluating transport properties (de los Santos and Franzese 2009, Franzese and de los Santos 2009). The results have offered a new interpretation of the diffusion anomaly (de los Santos and Franzese 2011, 2012, Franzese et al. 2010).

The FS model is defined as follows. For $N$ water molecules distributed in a volume $V$, we partition $V$ into $N$ equal cells, each with volume $v \equiv V/N$ that depends on $P$ and $T$. If $v_0$ is the van der Waals volume of a water molecule, then $v \geq v_0$, and $v_0/v$ is the cell density when there are no HBs. For example, near the liquid-gas transition, the number of HBs is negligible and it is convenient to associate to each cell and its corresponding molecule $i \in [i, ..., N]$ an index $n_i$, with $n_i = 0$ if $v_0/v \leq 0.5$, and $n_i = 1$ otherwise. Because the two-states variable $n_i$ does not include the volume variation due to the HBs, we can consider it as a discretized *density field* due to the van der Waals interaction. However, as we will discuss in the following, the formation of HBs leads to density heterogeneities, encoded within the model in a discretized density field with many states and the same resolution as the cell's grid.

---
[1] A preliminary version of the model and its mean-field solution at zeroth order was proposed in 2000 (Fanzese et al. 2000).

The thermodynamic average $\langle n_i \rangle$ reminds the order parameters for the liquid-gas phase transition of the lattice-gas model. However, here is always $\langle n_i \rangle = n_i$, and the order parameter, as we will discuss, is a more complex function of the molecular configuration.

An essential feature of water is its ability to form HBs between neighboring molecules. The HB has a strong directional component due to the dipole-dipole interaction between the highly concentrated positive charge on each H and each of the two excess negative charges concentrated on the O of another water molecule. The features of the HB in water have been studied over the last decades with a variety of models and computational techniques, from *ab initio* calculations to classical molecular dynamics simulations. All these studies emphasize that, at each pressure and temperature, the HB has a strong directional (covalent) component (Galkina et al. 2017) that qualitatively can be described by the $\widehat{OOH}$ angle of the O–H···O bond (see, for example, Ceriotti et al. 2013; Schran and Marx 2019). Recent calculations suggest that $\widehat{OOH} < 24°$ with ~50% probability at 250 K (Schran and Marx 2019). However, consistent with Debye-Waller factors estimates (Teixeira and Bellissent-Funel 1990), previous calculations (Ceriotti et al. 2013) showed that $\widehat{OOH} < 30°$ can be associated to a HB, confirming a previous assumption (Luzar and Chandler 1996). Hence, the HB must have $-30° < \widehat{OOH} < 30°$, i.e., only 1/6 of the entire range of possible values [0,360°] of the $\widehat{OOH}$ is associated with a bonded state.

In the FS model, this condition is accounted for by introducing a *bonding variable* $\sigma_{ij} = 1, ..., q$ for each molecule *i* facing the n.n. molecule *j*. By definition a HB between *i* facing and *j* is possible only if $\delta_{\sigma_{ij},\sigma_{ji}} = 1$, with $\delta_{ab} = 1$ if *a=b*, 0 otherwise, where the *bonding variables* have *q*=6 possible states. Therefore, only 1/6 of all the possible $(\sigma_{ij}, \sigma_{ji})$ configurations correspond to a HB.

Furthermore, calculations show that the larger the O-O distance, *r*, between the two molecules, the smaller is the HB strength (see, for example, Huš and Urbic 2012) in a such a way that it is possible to define a limiting value, $r_{max}$, above which the HB can be considered broken. Different authors have proposed different values for $r_{max}$, ranging from 3.1 Å (Schran and Marx 2019) to 3.5 Å (Luzar and Chandler 1996) in bulk water.

If we use as reasonable condition $r_{max} \simeq 3.65$ Å, considering that the van der Waals diameter of a water molecule is $r_0 \simeq 2.9$ Å, we find that, if $r > r_{max}$, then $r_0^3/r^3 \equiv v_0/v < 0.5$, i.e., $n_i = 0$ in the FS model. Hence, two nearest neighbors (n.n.) molecules *i* and *j* can form a HB if $n_i n_j = 1$. Because by definition is $n_i = n_j$, $\forall i \neq j$, then $n_i n_j = n_i = n_j = 1$, and $n_i$ can be considered as a *bonding index*.

Therefore, in the FS model, the geometrical definition of the HB is granted when two n.n. molecules 1) have the facing *bonding variables* in the same state, and 2) their *bonding indices* are set to 1. As a consequence, the number of HBs is given by

$$N_{HB} = \sum_{\langle i,j \rangle} n_i n_j \delta_{\sigma_{ij},\sigma_{ji}}, \qquad (1)$$

where the sum is performed over n.n. water molecules.

FS, following previous works (Sastry et al. 1996), make the reasonable assumption that the formation of HBs is the primary source of local density fluctuations. A water molecule fully bonded to its hydration shell, formed by other four water molecules in a tetrahedral configuration, occupies approximately the same volume of a hydrated water molecule with no HBs and larger coordination number (Soper and Ricci 2000). Therefore, to each HB we can associate a proper volume $v_{HB}$ given by 1/4 of the proper volume fluctuation in the hydration shell. A reasonable choice for this parameter is $v_{HB}/v_0 = 0.5$, equal to the average volume increase between (high density) ices VI and VIII and (tetrahedral, low density) ice Ih. Hence, the total volume $V_{tot}$ occupied by the system increases linearly with the number of HBs, $N_{HB}$, i.e.,

$$V_{tot} \equiv V + N_{HB} v_{HB}. \tag{1}$$

While $V$ is the volume without HBs, used for partitioning the system, and is homogeneously distributed among the $N$ water molecules, $V_{tot}$ includes the local heterogeneities in the density field due to the HBs.

The FS Hamiltonian is by definition

$$\mathcal{H} \equiv \mathcal{H}_{vdW} + \mathcal{H}_{HB} + \mathcal{H}_{Coop}, \tag{2}$$

where the first term accounts for the van der Waals (dispersive) attraction and hard core (electron) repulsion between water molecules and is

$$\mathcal{H}_{vdW} \equiv \sum_{i,j} U(r_{ij}), \tag{3}$$

summed over all the water molecules $i$ and $j$ at O-O distance $r_{ij}$, with

$$U(r) \equiv \begin{cases} \infty & \text{if } r \leq r_0 \\ 4\epsilon\left[\left(\frac{r_0}{r}\right)^{12} - \left(\frac{r_0}{r}\right)^{6}\right] & \text{if } r_0 \leq r \leq 25\, r_0 \\ 0 & \text{if } r > 25\, r_0 \end{cases} \tag{4}$$

a (double) truncated Lennard-Jones (LJ) interaction with $\epsilon \equiv 5.8$ kJ/mol, close to the estimate based on isoelectronic molecules at optimal separation $\simeq 5.5$ kJ/mol (Henry 2002).

The LJ potential is a function of the continuous variable $r$ and it is truncated at large distance and at short distance for numerical efficiency. Our previous analysis (de los Santos and Franzese 2011) shows that both truncations, introduced to simplify the implementation of the model, do not affect the results. This interaction regulates the liquid-gas phase transition and dominates above the liquid-gas spinodal temperature.

Because the formation of HBs, does not affect the distance $r$ between a molecule and those in its hydration shell (Soper and Ricci 2000), the van der Waals interaction is not affected by the HBs. This observation is crucial to state that the FS model is not mean field[2].

The second term is proportional to $N_{HB}$ and accounts for the additive (two-bodies) component of the HB,

$$\mathcal{H}_{HB} = -J\, N_{HB}, \tag{5}$$

with an energy decrease for each HB given by $J = 0.5 \times 4\epsilon \simeq 11$ kJ/mol, close to the estimate from the optimal HB energy and a HB cluster analysis (Stokely et al. 2010a).

Each new HB leads to an entropy decrease equal to $-k_B \ln 6$ ($k_B$ is the Boltzmann constant), due to the selection of one of the six possible states for the bonding variable. This entropy loss for increasing $N_{HB}$ is consistent with the anticorrelation between entropy and volume that, for the Clausius-Clapeyron relation, is responsible for the negative slope of the melting line of water near ambient pressure. The total number of accessible states for a free water molecule with four bonding variables is $q^4 = 1296$, for $q=6$, and for a system of $N$ molecules at high temperature is $1296^N$.

---

[2] We acknowledge the late Professor David Chandler for discussing this point.

The third term of the Hamiltonian accounts for the HB cooperativity due to many-body correlations (Barnes et al. 1979). The many-body interactions lead to the local order in the hydration shell (Soper and Ricci 2000) and are a consequence of the quantum nature of the HB (Barnes et al. 1979, Hernández de la Peña and Kusalik 2005, Ludwig 2001). This term is modeled in classical atomistic potentials with a long-range dipolar interaction, as for example in (Dang and Chang 1997). In the FS Hamiltonian, it is modeled as a five-bodies term in which each molecule $i$ is interacting with its first coordination shell

$$\mathcal{H}_{Coop} = -J_\sigma \sum_i n_i \sum_{(k,l)_i} \delta_{\sigma_{ik},\sigma_{il}}, \qquad (6)$$

where $(l,k)_i$ indicates each of the six different pairs of the four variables $\sigma_{ij}$ of the molecule $i$ and $J_\sigma$ is the extra energy-gain provided by each cooperative HB. A direct experimental evaluation of such a term is not available, however it can be estimated by attributing to it (Heggie et al. 1996) the 3 kJ/mol increase in strength of the HBs in ice Ih with respect to liquid water (Eisenberg and Kauzmann 1969). This value is consistent by order of magnitude with the 6 kJ/mol estimated for $D_2O$ (Shimoaka et al. 2012), for which the effect is expected to be stronger. Considering that each HB participates to three terms in Eq. (6), we can estimate the value of $J_\sigma$ to be ~ 1.0 kJ/mol. Because $J_\sigma \ll J$, this term is relevant only when $N_{HB} \gg 1$. This asymmetry between the two components of the HB interaction is necessary for the model to represent water.

In the *NPT* ensemble the partition function of the system is

$$Z(T,P) \equiv \sum_{\{\sigma\}\{v\}} e^{-H/k_B T}, \qquad (7)$$

where the sum is over all the possible configurations of bonding variables $\{\sigma\}$ and cell volumes $\{v\}$, and

$$H \equiv \mathcal{H} + P V_{tot} \equiv \mathcal{H}_{vdW} - J_{eff} \sum_{\langle i,j \rangle} n_i n_j \delta_{\sigma_{ij},\sigma_{ji}} + \mathcal{H}_{Coop} + PNv \qquad (8)$$

is the enthalpy, with $J_{eff} \equiv J - P v_{HB}$, the effective interaction between $\sigma$-variables of n.n. molecules, that depends on $P$.

Eq. (7) can be rewritten as

$$Z(T,P) = \sum_{\{v\}} e^{-[U(v)+PNv]/k_B T} \times Z_{\{\sigma\}} \qquad (9)$$

where

$$Z_{\{\sigma\}} \equiv \sum_{\{\sigma\}} e^{(J_{eff}/k_B T) \sum_{\langle i,j \rangle} n_i n_j \delta_{\sigma_{ij},\sigma_{ji}}} \times e^{\left(\frac{J_\sigma}{k_B T}\right) \sum_i n_i \sum_{(k,l)} \delta_{\sigma_{ik},\sigma_{il}}}$$

$$= \sum_{\{\sigma\}} \prod_{\langle i,j \rangle} \left[ 1 + \left( e^{\left(\frac{J_{eff}}{k_B T}\right)} - 1 \right) n_i n_j \delta_{\sigma_{ij},\sigma_{ji}} \right] \qquad (10)$$

$$\times \prod_{i=1}^{N} \prod_{(k,l)_i} \left[ 1 + \left( e^{\left(\frac{J_\sigma}{k_B T}\right)} - 1 \right) n_i \delta_{\sigma_{ik},\sigma_{il}} \right].$$

Here $\prod_{\langle i,j \rangle}$ runs over all the n.n. molecules, $\prod_{i=1}^{N}$ runs over all molecules and $\prod_{(k,l)_i}$ extends over all the six pairs of the bonding variables of a specific molecule $i$.

Due to its simplicity, it is possible to perform a thorough analysis of how the macroscopic properties of the model depend on its limited number of parameters, each describing a molecular mechanism, both with theoretical calculations with the *cavity method* and with MC simulations (Bianco et al. 2013a, Franzese et al. 2010, Franzese and Stanley 2002a,b, 2007, Franzese et al. 2008, Kumar et al. 2008c, Mazza et al. 2011, 2012, 2009, Stanley et al. 2010, 2011, 2009, Stokely et al. 2010a,b). The model, initially defined for monolayers with height $h = 0.5$ nm and $v \equiv hr^2$, has recently been extended to bulk water (Coronas et al. 2016).

## 4. Dynamic behavior of the FS water model

### 4.1 Two dynamic crossovers in FS water

Data on protein hydration water and confined water show the presence of a dynamical crossover from a non-Arrhenius to an Arrhenius regime (Franzese et al. 2008) at $\sim 220$ K. For example, this crossover is found in the translational correlation time of water hydrating lysozyme proteins (Chen et al. 2006b) or in the structural relaxation time of water confined in silica pores (Faraone et al. 2004, Liu et al. 2005). Also, simulations of the TIP5P water model show a dynamic crossover from a non-Arrhenius to an Arrhenius regime in the diffusivity of water hydrating lysozyme and DNA (Kumar et al. 2006b). As this crossover takes place at much higher temperatures than $T_G$, the so-called *glass transition temperature*, Kumar et al. discard any relation with the glass state. According to this numerical and experimental evidence, a possible hypothesis, among others (Cerveny et al. 2006, Swenson 2006), for the origin of this crossover is the local rearrangement of the HB network at low temperatures (Chen et al. 2006, Kumar et al. 2008c).

In particular, MC simulations of the FS model display a dynamic crossover from a non-Arrhenius to an Arrhenius regime that is a consequence of the local rearrangement of the water HBs (Kumar et al. 2008b,c). Kumar et al., by mean-field calculations and MC simulations of the FS water monolayer, estimate the orientational correlation time, i.e., the relaxation time $\tau$ of the autocorrelation $C_S(t) \equiv < S_i(t) S_i(0) > / (S_i^2)$, where $S_i \equiv \sum_j \sigma_{ij}/4$, for a FS monolayer. The physical meaning of $S_i$ is the total bond ordering of the $i$-th water molecule, and $\tau$ is defined as the time at which $C_S(t)$ decays by a factor $1/e$. Kumar *et al.* find a non-Arrhenius regime at high-$T$ upon cooling at constant pressure, where $\tau$ can be fitted with the Vogel-Fulcher-Tamman (VTF) function

$$\tau^{\text{VTF}} \equiv \tau_0^{\text{VTF}} \exp\left[\frac{T_1}{T - T_0}\right], \tag{11}$$

with $\tau_0^{\text{VTF}}$, $T_0$, and $T_1$ fitting parameters. At low-$T$, $\tau$ displays Arrhenius behavior, $\tau = \tau_0 \exp[E_A/k_B T]$, where $\tau_0$ is the limiting time at high-$T$, and $E_A$ is the $T$-independent activation energy. The crossover occurs at the same temperature where the specific heat displays a maximum. The authors discuss that they interpret this dynamic crossover as an effect of the breaking and reorientation of HBs, leading to a more tetrahedrally structure for the HB network at low-$T$.

In (Kumar et al. 2008c), the authors also investigate how the cooperative term affects the dynamics. At constant $J/\epsilon = 0.5$, they compare the results for $J_\sigma/\epsilon = 0.05$ and $J_\sigma = 0$, corresponding to the LLCP scenario and the SF scenarios, respectively, as we will discuss in the following. In both scenarios, they find that (i) the crossover time $\tau_C$ is approximately $P$-independent, (ii) the Arrhenius activation energy $E_A(P)$ decreases upon increasing $P$ and (iii) the temperature $T_A(P)$, at which $\tau$ reaches a fixed macroscopic time $\tau_A \geq \tau_C$, decreases upon increasing $P$. Furthermore, they show that (iv) $E_A/(k_B T_A)$ increases upon pressurization in the LLCP scenario, but it remains constant in the SF scenario.

These new predictions have been tested in experiments in a protein hydration layer (Chu et al. 2009, Franzese et al. 2008). In particular, quasi-elastic neutron scattering (QENS) (Chu et al. 2009, Franzese et al. 2008) verifies that the predictions (i)-(iii) are correct for a water monolayer hydrating lysozyme. Nevertheless, the resolution of these experiments does not allow to settle, on the base of the prediction (iv), which among the LLCP and SF scenario is satisfied by proteins hydration water (Franzese et al. 2008).

Further research on water monolayers at lower $T$, surprisingly, has shown the presence of not one, but two dynamic crossovers (Mazza et al. 2011). Mazza *et al.* measure the dielectric relaxation time of water protons $\tau_{WP}$, as it is sensitive to breaking and formation of HBs (Peyrard 2001), in a water monolayer hydrating lysozyme protein, and compare it with MC simulation of a FS water monolayer. In both cases, the authors find two crossovers at ambient pressure. The first is at $T \sim 252$ K, as previously reported by Kumar and coworkers (Kumar et al. 2008b). The second is at $T \sim 181$ K.

They find that the crossover at $T \sim 252$ K is between two VTF behaviors. They interpret this result as a change in the diffusion of water protons, between a high-$T$ diffusive regime and a low-$T$ sub-diffusive regime.

The crossover at $T \sim 181$ K, the crossover is between a VTF and an Arrhenius regime, corresponding to the rearrangement of the HB network structure. The experimental measurements of $\tau_{WP}$ compare well with FS water calculations of the $S_i$ relaxation time from MC simulations $\tau_{MC}$. Mazza and coworkers consider the autocorrelation function

$$C_M(t) \equiv \frac{1}{N} \sum_i \frac{\langle S_i(t_0+t) S_i(t_0) \rangle - \langle S_i \rangle^2}{\langle S_i^2 \rangle - \langle S_i \rangle^2}, \qquad (12)$$

which decays to 0 as $t \to \infty$. By definition, is $C_M(0) = 1$. Following (Kumar et al. 2008c), $\tau_{MC}$ is by definition such that $C_M(\tau_{MC}) = 1/e$. They find two dynamic crossovers and relate them to the presence of two specific heat maxima: the high-$T$ weak maximum and the low-$T$ strong maximum. At high-$T$, the $C_P$ weak maximum occurs when the fluctuations of $N_{HB}$ are maximum. At low-$T$, the $C_P$ strong maximum is due to the maximum in the fluctuations of the cooperative term (Eq. 6) of the Hamiltonian. Hence, the high-$T$ crossover is associated with the formation of the HB network. Instead, the low-$T$ crossover is due to the rearrangement of the HBs in an ordered structure in the monolayer.

### 4.2 Effect of pressure and temperature on the dynamics

Protein hydration water undergoes a liquid-glass transition (LGT) with the glass state characterized by a huge increment of viscosity and the freezing of long-range translational diffusion (Doster 2010). The macroscopic structural arrest of the glass state emerges from the slowing down of nearest neighbors HBs dynamics. Doster and coworkers perform neutron scattering experiments on myoglobin at hydration level $h = 0.35 g_{H_2O}/g$, as this technique allows for monitoring displacements at the microscopic scale. They measure the incoherent intermediate scattering functions $I(q,t)$ at $T = 320$ K (above the LTG but relevant for translational diffusion) and for scattering vector $0.4 \leq q/\text{Å}^{-1} \leq 2$ (Settles and Doster 1996). They find a two-step time-decay in $I(q,t)$ that, at high-$q$ and long times $t$, can be fitted to a stretched exponential

$$C(t) = C_0 \exp[-(t/\tau_0)^\beta], \tag{13}$$

where $\tau_0$ is the correlation time, $0 < \beta \leq 1$ is the stretched exponent and $C_0$ is a normalization factor. For large $q$, their results show that $0.3 \leq \beta \leq 0.4$. When they measure $I(q, t)$ at constant q = 1.8 Å$^{-1}$ and $180 \leq T/K \leq 320$ as the temperature decreases, the two-step decay turns into a plateau, which leads to a relaxation time that exceeds the observation time (Doster 2010).

Calculations from MC simulations of the FS model are consistent with these two experimental results. In particular, to study the microscopic origins of the complex dynamics of water on low-hydrated proteins, de los Santos and Franzese consider a monolayer of FS water adsorbed on a generic inert substrate at 75% hydration (de los Santos and Franzese 2011, 2012, Franzese and de los Santos 2009). At such hydration level, adsorbed water molecules are restricted to diffuse on a surface geometry with an up to four coordination number. They calculate $C_M(t)$ at different $T$ and $P$ (Franzese and de los Santos 2009) finding that at high pressure, $P \geq 1\ \epsilon/v_0$, the correlation function decays exponentially for any $T$ allowing the system to equilibrate easily. At these pressures, the HB network is inhibited inducing rapid dewetting and large dry cavities with decreasing temperature.

At lower pressure, $P = 0.7\ \epsilon/v_0$, and low $T$, the behavior of $C_M(t)$ can be fitted with a stretched exponential function, with no strong increase of the correlation time as $T$ decreases (Fig.1). The authors associate this behavior with (i) the rapid ordering of the HBs that generates heterogeneities and (ii) with the lack of a single timescale due to the vicinity of the liquid-liquid critical point, as we will discuss in the next section.

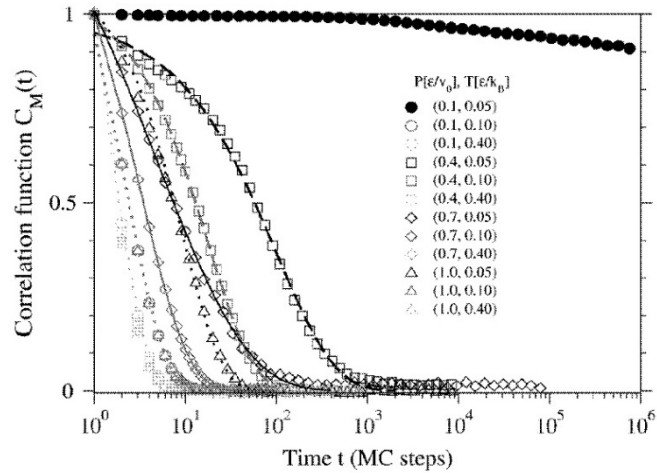

Fig. 1: The correlation function $C_M(t)$ for pressures $P \leq 1.0\ \epsilon/v_0$ and temperatures $T \leq 0.4\ \epsilon/k_B$. $C_M(t)$ decays exponentially (dotted lines) at $P = 1.0\ \epsilon/v_0$ for all $T$ (triangles) and at $P < 1.0\ \epsilon/v_0$ for any $T \geq 0.4\ \epsilon/k_B$ (not shown). For $T \leq 0.1\ \epsilon/k_B$ and $P = 0.7\ \epsilon/v_0$ (diamonds, continuous line) or $P = 0.4\ \epsilon/v_0$ (squares, dashed line), $C_M(t)$ can be described by a stretched exponential, with an exponent $\beta$ that decreases as $T$ is lowered. At low pressure $P = 0.1\ \epsilon/v_0$ (circles), $C_M(T)$ is exponential for $T \geq 0.1\ \epsilon/k_B$ and non-exponential at $T = 0.05\ \epsilon/k_B$ (solid circle). (Figure reprinted from (Franzese and de los Santos 2009) "Dynamically slow processes in supercooled water confined between hydrophobic plates". Copyright (2009) IOP Publishing. Reproduced with permission. All rights reserved).

At even lower pressures, the gradual formation of the HB network, starting at higher $T$, is responsible for the dynamic slowing down as $T$ decreases and for the dynamical arrest at $(P, T) = (0.1\ \epsilon/v_0, 0.05\ \epsilon/k_B)$, with an increase in $\tau_0$ of more than four orders of magnitude, as in a glass. Under these conditions, the dewetting process is strikingly different, with the formation of many small cavities.

Comparison between FS results and experiments (Doster 2010, Settles and Doster 1996) show that the complex dynamic behavior of protein hydration water at low hydration can be well reproduced by solely taking into account the dynamics of the HB network. The LGT emerges as a consequence of the slowing of the HBs dynamics by decreasing $T$. At extremely low $T$, this results in a dynamic arrest of the system.

### 4.3 The diffusion anomaly

Experiments and simulations of the diffusion of confined water show controversial results. For water confined in NaX and NaA zeolites and for $T$ between 310 K and 260 K, experiments observe a reduction of two orders of magnitude of the translational diffusion coefficient $D$ respect to the bulk case (Kamitakahara and Wada 2008). Other results show that $D$ decreases upon increasing the confinement in either hydrophilic (Takahara et al. 1999) or hydrophobic conditions (Naguib et al. 2004, Takahara et al. 1999). However, for confined water in carbon nanotubes with a diameter smaller than 2 nm, the experiments find extremely fast mass transport (Holt et al. 2006). Furthermore, water confined in smooth graphene capillaries shows a fast flow (of ∼1 m/s) that is enhanced if the height of the channel can accommodate only a few water layers (Radha et al. 2016). The authors associate the fast flow to the great capillary pressure and relate its enhancement to the increased structural order of nanoconfined water.

Several models, e.g. (Netz et al. 2001, Szortyka and Barbosa 2007), can reproduce numerically the diffusion anomaly, but they display a variety of different results still controversial. Classical molecular dynamics simulations of SPC/E water in hydrophilic MCM-41 or Vycor show that the mobility of water molecules decreases as the hydration level is lowered (Gallo et al. 2010). This is interpreted as due to the greater proportion of molecules bonded to the surface at low hydration.

A diffusion decrease, by two orders of magnitude compared to the bulk, is found in simulations of TIP5P water between hydrophobic, smooth, planar plates. In particular, the diffusion is anomalous in the direction parallel to the walls, while it is not for the orthogonal direction (Han et al. 2008, Kumar et al. 2005). However, first-principle molecular dynamics simulations of SPC/E water confined in graphene sheets and carbon nanotubes show a faster diffusion compared to the bulk case (Cicero et al. 2008), possibly due to weaker H-bonding at the interface.

To shed light on this controversy, de los Santos and Franzese perform MC simulations of a water monolayer confined in a smooth slit pore at partial hydration. Their results describe the origins of the diffusion anomaly (de los Santos and Franzese 2011, 2012) in terms of *cooperative rearranging regions* (CRR) of water. They calculate the diffusion coefficient $D_{//}$ parallel to the walls using Einstein's formula

$$D_{||} = \lim_{t \to \infty} \frac{\langle |\vec{r}_i(t+t_0) - \vec{r}_i(t)|^2 \rangle}{4t}, \tag{14}$$

where $\vec{r}_i$ is the projection of the position of molecule $i$ onto the plates. The average $\langle \cdot \rangle$ is over all molecules and different times $t_0$. The analysis of $D_{//}$ shows the presence of maxima and minima along isotherms at high temperature (Fig. 2).

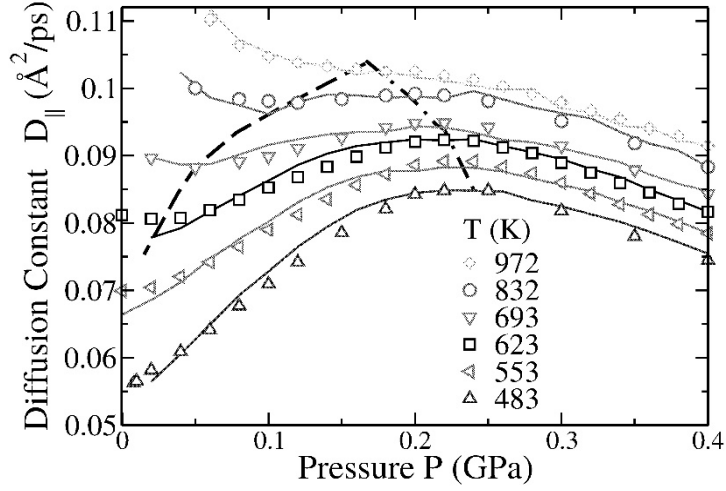

Fig, 2: Diffusion coefficient $D_{//}$ from MC simulations (symbols) as a function of pressure along isotherms. For $T <972$ K, $D_{//}$ has maxima (dotted-dashed line) and minima (dashed line). Solid lines are from $W_{v,\mu}$ calculations (defined in the text). (Reprinted figure with permission from (de los Santos and Franzese 2012). Copyright (2012) by the American Physical Society).

The authors describe the anomaly in terms of the joint probability

$$W_{v,\mu}(P,T) \equiv \mathcal{P}_F^v \mu \mathcal{P}_b \frac{1}{Z} \exp[-H(T,P)/k_B T] \tag{15}$$

of finding $v$ molecules with a free cell available for diffusion within a region with a number $\mu \mathcal{P}_b$ of HBs, where $Z$ is the partition function, $\mathcal{P}_F$ and $\mathcal{P}_b$ are the probability for each cell to have a free n.n. and the probability for each HB to be formed, respectively, with $\mathcal{P}_F \equiv <n_F>/4$, $\mathcal{P}_b \equiv <n_{HB}>/4$, $<n_F>$ the average number of free n.n. cells per molecule and $<n_{HB}>$ the average number of HBs per molecule.

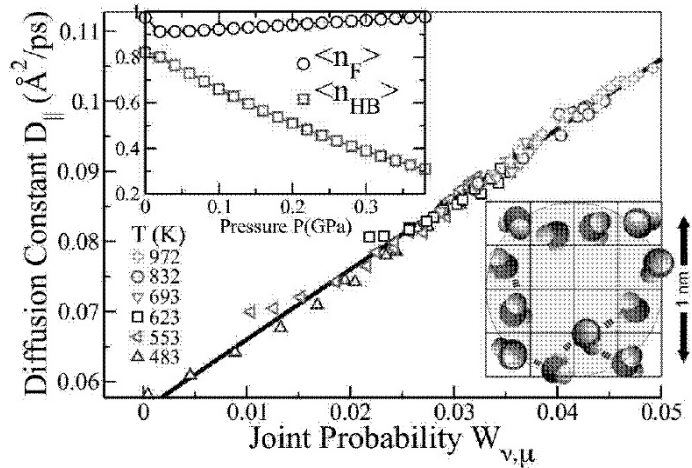

Fig. 3: Upper inset: Average number of free n.n. cells around a molecule $<n_F>$ and average number of HBs formed by a molecule $<n_{HB}>$ at $T = 693$ K. There is a discontinuity in $<n_F>$ at low $P$ corresponding to the gas-liquid phase transition. At high $P$ they are both monotonic. Main panel: Linear dependency of $D_{//}$ vs. $W_{v,\mu}$ along the isotherms represented in Fig. 2. Lower inset: Example of a CRR (shaded) of about 1 nm size, with $v= 12$ molecules and $\mu <n_{HB}>/4 = 5$ HBs. (Reprinted figure with permission from (de los Santos and Franzese 2012). Copyright (2012) by the American Physical Society).

The authors find that $D_{//}$ is proportional to $W_{v,\mu}$, implying that the diffusion is dominated by the cooperativity of water (Fig. 3 main panel). The resulting value for $v = 12.5 \pm 0.5$ suggests that diffusion requires a CRR that reaches ~3.5 molecules (Fig. 3 lower inset). Thus, the FS model clarifies that the diffusion anomaly, at constant $T$ by increasing $P$, originates from the competition between, on the one hand, the increase of free volume $<n_F>$ and the decrease of the energy cost for a molecule to move due to the reduction of $<n_{HB}>$ (Fig. 3 upper inset), and, on the other hand, the decrease of free volume due to the increase of density. The diffusion is favored by pressurizing at constant $T$ until the HB formation is unfavorable for enthalpic reasons, giving origin to the $D_{//}$ maxima. The diffusion coefficient $D_{//}$ correlates to the phase diagram of FS water (Fig. 4). The loci of $D^{max}$ and $D^{min}$ along isotherms lie between the temperature of maximum density (TMD) line and the liquid-gas spinodal. The constant $D_{//}$ lines resemble the melting line of bulk water. In the deeply supercooled region, there is a subdiffusive regime due to the increment of the relaxation time of the HB network, which at low $P$ and $T$ leads to the dynamical arrest and to the amorphous glassy water (Handle et al. 2017).

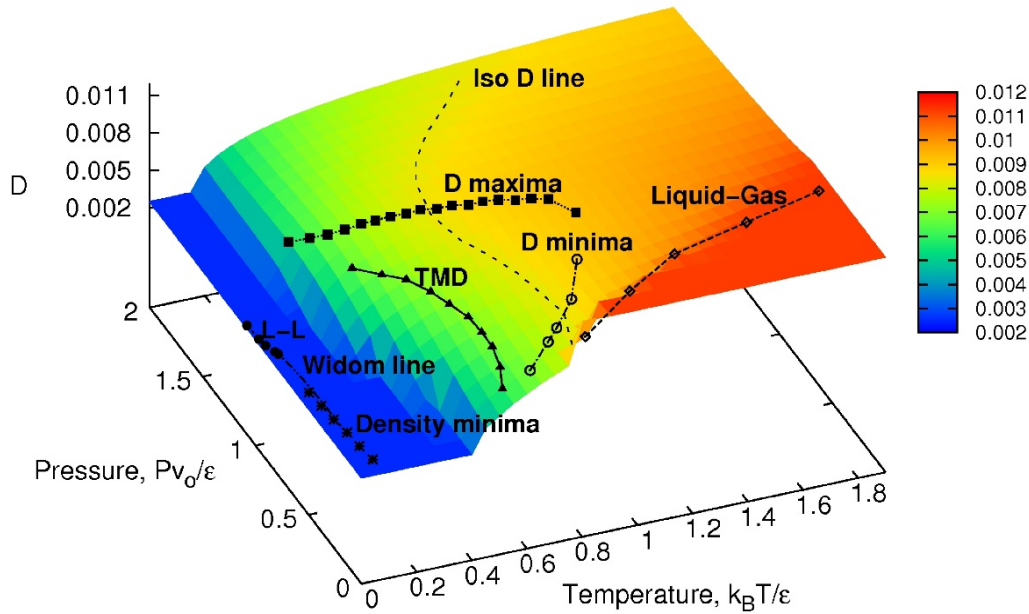

Fig. 4: Phase diagram of a water monolayer nanoconfined between hydrophobic plates. Z-axis and color scale represent de diffusion constant $D_{||}$ for values $0.12 \geq D_{||}$ (Å/ps$^2$)$\geq 0.03$. At high $T$ and low $P$ there is the liquid-gas phase transition line (open diamonds) ending in the liquid-gas critical point (diamond symbol at highest $T$), as discussed in section 5. The loci of isothermal $D_{||}$ extrema, $D^{max}$ (solid squares), and minima, $D^{min}$ (open circles), envelope the TMD line (solid triangles). Loci at constant $D_{||}$ (e. g., the dashed line marked as "Iso-$D$") resemble in their reentrant behavior the water melting line. (Reprinted with permission from (de los Santos and Franzese 2011). Copyright (2011) American Chemical Society).

## 5. FS-water phase diagram

The phase diagram of the FS model has been extensively studied in the case of a water monolayer by analytic (Franzese and Stanley 2002a,b, Franzese et al. 2000) and numerical methods (Franzese et al. 2003, de los Santos and Franzese 2009). By changing the model's parameters, Stokely et al. reveal the relations among the different scenarios for the anomalies of water (Stokely et al. 2010a).

The $T$-$P$ plane (Fig. 5) has several features (Bianco and Franzese 2014).
- The isobaric TMD line
    - (i) at high $P$, has a negative slope, as in the experiments, while
    - (ii) at negative $P$, it has a positive slope and asymptotically approaches the liquid-gas (LG) spinodal.
    - (iii) At the turning point, it crosses the line of minima of the isothermal compressibility along isobars, $K_T^{min}(T)$ (Fig. 5a).
    - (iv) It avoids crossing the LG spinodal line at low $T$ and turns into an isobaric line of the temperature of minimum density with a negative slope, as suggested by experiments (Mallamace et al. 2007a).
    - (v) Its low-$T$ turning point occurs where the line of (weak) minima of $\alpha_P$ along isobars, $\alpha_P^{wmin}(T)$, crosses it (Fig. 5c).
    - (vi) This point is also where the line of specific heat (weak) maxima along isotherms, $C_P^{wMax}(P)$, is aiming for (Fig. 5b).
- (vii) The locus of (weak) maxima of compressibility along isobars, $K_T^{wMax}(T)$ turns, at its minimum $P$, into the locus of minima of $K_T$ along isobars, $K_T^{min}(T)$ (Fig. 5a).
- (viii) The resulting line of extrema of isobaric $K_T$ coincides within the error bar (not shown) with the locus of (weak) minima of $\alpha_P$ along isotherms, $\alpha_P^{wmin}(P)$ (Fig. 5c).

All the findings (i)-(viii) are consistent with thermodynamic relations and atomistic models (Poole et al. 2005), confirming the correctness of the numerical calculations performed in Ref. (Bianco and Franzese 2014).

Thanks to its mapping onto a percolation formulation (Bianco and Franzese 2019), the FS water can be easily simulated adopting a fast (cluster) MC algorithms (Mazza et al. 2009) that allows us to explore in detail the deeply supercooled states (Franzese and Stanley 2007) and the high-pressure region (Franzese et al. 2010, 2008, Stokely et al. 2010b). In particular, in the deeply supercooled region, Mazza et al. (2011, 2012) and Bianco et al. (Bianco and Franzese 2014, Bianco et al. 2013a) reveal the presence of strong extrema, for the response functions $C_P$, $K_T$ and $\alpha_P$, occurring at temperatures lower than those for the weak extrema (Fig. 5). While the weak extrema occur both in the FS model and in atomistic water models, the strong extrema have been found only in the FS model for the monolayer. Their presence in bulk models, both FS and atomistic, for water is still under investigation. It is important to observe here that the FS model predicts the strong extrema in a region in which the atomistic models freeze into a glassy state (Kesselring et al. 2012, 2013, Palmer et al. 2014).

Mazza and coauthors (Mazza et al. 2011) show that the two (weak and strong) maxima of the response functions correspond to the two dynamic crossovers, discussed in section 4.1, found in experiments and the FS model for a hydration monolayer. In this way, they establish a connection between thermodynamics and dynamics that extends up to moderate supercooling at low $P$.

As initially observed by Franzese and coworkers (Bianco and Franzese 2014, Franzese and Stanley 2007, Franzese et al. 2008, Kumar et al. 2008b,c, Mazza et al. 2012, Stokely et al. 2010b), all the loci of the extrema of response functions converge toward a point (large circle with label A in Fig. 5). At this point A, the extrema reach their maximum values, as expected at a critical point in a finite-size system. This point A corresponds to the LLCP, as shown by analyzing the fluctuations around it (Mazza et al. 2011, Mazza et al. 2012). In particular, as described with more detail in the next section, a finite-size scaling analysis shows that the LLCP in the FS monolayer belongs to the 2D Ising universality class (Bianco and Franzese 2014).

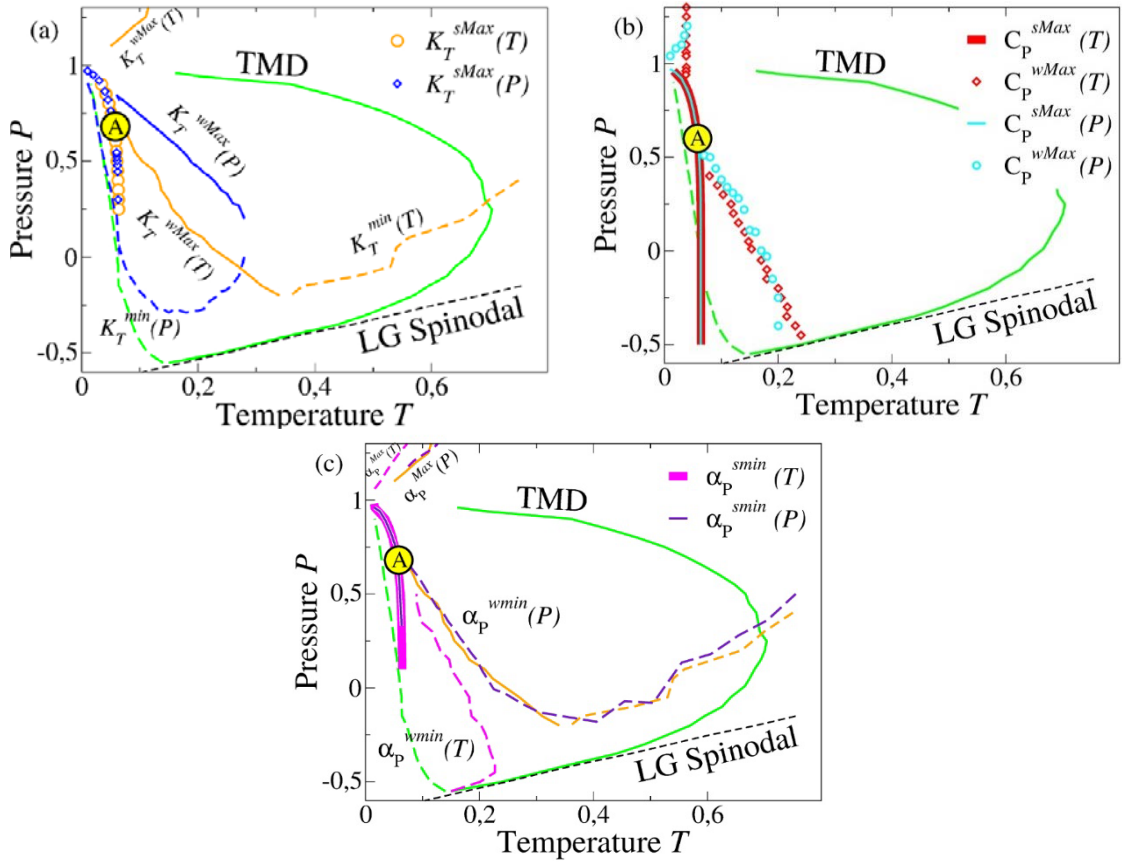

Fig. 5: Phase diagram of a water monolayer. In all panels, the black dashed line represents the LG spinodal, while all the colored dashed lines turning into solid lines of the same color represent loci of minima and maxima, respectively, of different quantities: e.g., in green the temperature of minimum density and the TMD. Each panel focuses on one response function: (a) $K_T$ along isobars (isotherms) in orange (blue), with dashed (continuous) lines for weak minima (maxima) and symbols for strong maxima; (b) $C_P$ along isobars (isotherms) in red (turquoise), with lines (symbols) for strong (weak) maxima; (c) $\alpha_P$ along isobars (isotherms) in magenta (violet), with dashed (continuous) lines for weak (strong) minima. All the loci of extrema of response functions converge toward the large circle with label A corresponding to the LLCP. (Reprinted with permission from (Bianco and Franzese 2014)).

This result is consistent with those for ST2 water (Kesselring et al. 2012, Palmer et al. 2014, Smallenburg and Sciortino 2015), TIP4P/2005 water (Abascal and Vega 2010, González et al. 2016), but not with those for the three-body interactions mW model (Limmer and Chandler 2013). Works by Anisimov and coauthors show that the non-ideality of mixing of the mW case is entropy-driven, instead of energy-driven as in the models with the LLCP, being not strong enough to induce the liquid-liquid phase separation (Holten et al. 2013, Singh et al. 2016).

Extrapolations from the experiments show large fluctuations for the response functions (Fuentevilla and Anisimov 2006, Holten and Anisimov 2012, Kim et al. 2017) as calculated in the FS-water, at variance with the atomistic models. This significant difference could be a consequence of the explicit many-body interactions included in the FS model, but not in the (non-polarizable) atomistic models. Further investigation is necessary on this difference.

The FS water monolayer presents the first-order LLPT, between two liquids with different energies and structures, at pressures above (and $T$ below) the LLCP. Other (atomistic) models adopting a two-state description of bulk water show the same property (Gallo et al. 2016, Holten et al.

2013, 2014, Russo and Tanaka 2014, Shi et al. 2018, Singh et al. 2016). In the FS model, the phase transition occurs between the HDL, with a disordered HB network and high energy, and the LDL, with an ordered HB network and low energy (Franzese et al. 2003).

### 5.1 The critical point analysis

Bianco and Franzese study the universality class of the LLCP (Bianco and Franzese 2014). To define the correct order parameter $M$ for the LLCP, they first analyze the free-energy landscape. For the LLCP in a finite-size system, the landscape has two equivalent minima, each with a basin of attraction, with different energy and density, corresponding to two coexisting HDL and LDL phases (Fig. 6.a). They find that a free energy barrier $\sim k_B T$ separates the two basins. Hence, as expected near a critical point, the system has enough thermal energy to cross the barrier between the two basins.

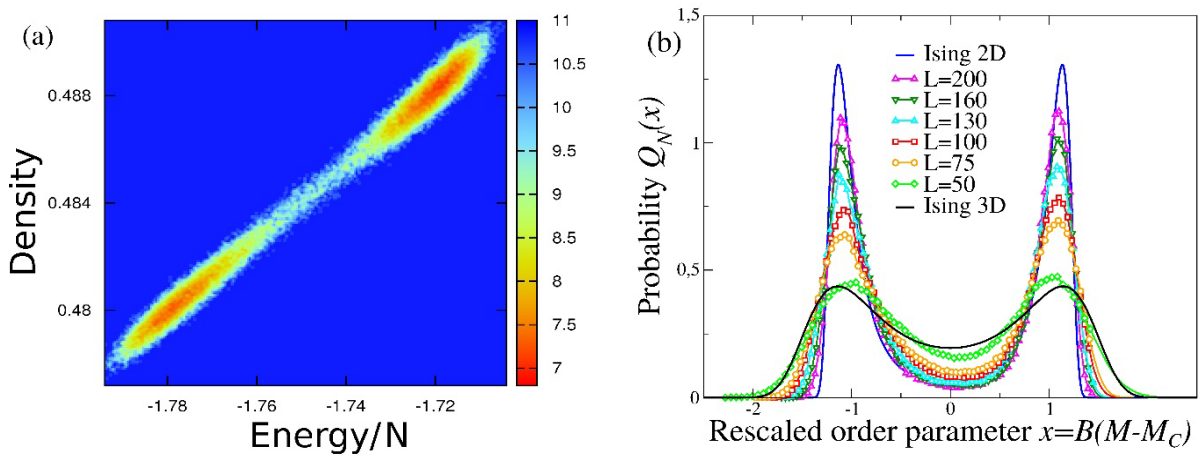

Fig. 6: The free-energy landscape in the vicinity of the LLCP. (a) The Gibbs free energy, $G$ in units of $k_B T$, is represented as heat-map in the energy-density plane, showing two basins separated by a barrier $\Delta G \sim k_B T$, as expected near a critical point. The phase transition is described by the order parameter $M$ given by the reaction coordinate (an imaginary straight line in the plane, not shown) between the two minima. (Reprinted with permission from Springer Nature Customer Service Centre GmbH: Springer Food Biophys. (Franzese and Bianco 2013), Copyright (2013)). (b) At the LLCP, as the size $L$ of the system increases, the $M$ probability distribution, $Q_N$, approximates the 2D-Ising universality-class (blue solid line). Here, $B$ is a scaling factor, and the black reference distribution holds for the 2D-Ising model. (Reprinted with permission from (Bianco and Franzese 2014)).

The order parameter $M$ is, by definition, the reaction coordinate between the two minima, i.e., the linear combination of energy and density $M \equiv \rho + sE$, as in the mixed-field approach (Bruce and Wilding 1992, Franzese and Bianco 2013, Wilding 1995). $M$ is related to the diagonal connecting the centers of the two wells (Fig. 6.a), as its probability distribution must be symmetric at the LLCP. By performing a finite-size scaling analysis at fixed sample thickness $h = 0.5$ nm, and by varying the number of water molecules from 2500 to 40000 at fixed density, Bianco and Franzese find that the FS monolayer in the thermodynamic limit displays a LLCP belonging to the 2D-Ising universality-class (Fig. 6.b).

The probability distribution of $M$ shows a crossover from 3D to 2D-Ising universality-class as the number of water molecules in the monolayer increases (Fig. 6.b). The system crosses from 3D to 2D behavior for $L/h > 50$, where $L$ is the system lateral size, while for normal liquids this crossover takes place for $L/h > 5$ (Liu et al. 2010). This can be interpreted as a consequence of the high cooperativity, and the low coordination number, of the water molecules. For small $L$, the strong cooperativity at the

LLCP increases the HBs fluctuations, resulting in a (3D-like) probability distribution for $M$ broader than the one in 2D. The water coordination number–four both in 3D and 2D–emphasizes the effect because it reduces the fluctuations differences between the two cases when the system is small (Bianco and Franzese 2014, Franzese and Bianco 2013).

### 5.2 The Widom line

The Widom line (WL) is defined as the locus of maxima of the statistical correlation length $\xi$, emanated from the critical point as the analytic continuation of the first-order transition line, and spanning into the supercritical region (Franzese and Stanley 2007, Holten et al. 2012, Kumar et al. 2007). The expression "Widom line" was first used by Stanley and coworkers in 2005 as the locus where the lines of the maxima for different response functions asymptotically converge approaching the critical point from the supercritical region (Xu et al. 2005b, Kumar et al. 2007, Franzese and Stanley 2007).

Close to the critical point, the thermodynamic response functions can be expressed as power-law functions of $\xi$. Since $\xi$ by definition is maximum along the WL, the loci of maxima of thermodynamic response functions converge toward the WL on approaching the critical point and, near $(T_c, P_c)$, they are often used as *proxies* for the Widom line (Abascal and Vega 2010, Simeoni et al. 2010).

Following Ref. (Bianco and Franzese 2014), it is possible to calculate the correlation length $\xi$ using the spatial correlation function $G(r)$, that, within the FS model, is defined as

$$G(r) \equiv \frac{1}{4N} \sum_{|\vec{r_i}-\vec{r_l}|=r} \left[ \langle \sigma_{ij}(\vec{r_i})\sigma_{lk}(\vec{r_l}) \rangle - \langle \sigma_{ij} \rangle^2 \right]. \tag{16}$$

Far from the critical region, $G(r)$ decays exponentially, and $\xi$ is by definition the characteristic length of the decay, $G(r) \sim e^{-r/\xi}$. By approaching the critical point, the correlation function can be written as $G(r) \sim e^{-r/\xi}/r^{d-2+\eta}$, being $d$ the system dimension and $\eta$ a (critical) positive exponent (Stanley 1971).

In the FS model for a water monolayer, the correlation length can be calculated along isobars down to deeply supercooled temperatures, showing that it exhibits a maximum, $\xi^{Max}$ at all the $P$ explored. The $T$–$P$ locus of $\xi^{Max}$ overlaps (i) with the LLPT at high $P$ and (ii) with the locus of strong maxima of the specific heat at lower $P$ (Mazza et al. 2011, 2012). This line, at low $P$, identifies the WL (Mazza et al. 2012, Bianco and Franzese 2014), with a large slope in the $T$–$P$ plane, consistent with extrapolations from experiments (Fuentevilla and Anisimov 2006, Taschin et al. 2013). This finding, however, is at variance with previous works identifying the WL line in supercooled water with the higher–$T$ weak maxima of the response function (Xu et al. 2005a, Kumar et al. 2007, Franzese and Stanley 2007, Franzese et al. 2008, Mallamace et al. 2008, Abascal and Vega 2010).

Nevertheless, the FS water monolayer is the only model with the LLCP where $\xi$-maxima have been found. Furthermore, it is the only model that can be equilibrated so deeply into the supercooled region. Therefore, the prediction of a WL below the temperatures accessible to other models or experiments is consistent with the experimental observation of an increasing isobaric $\xi$ for decreasing $T$ but no maxima (Huang et al. 2010).

In a recent work, Bianco and Franzese use a percolation approach (Coniglio and Klein 1980, Kasteleyn and Fortuin 1969) to identify the regions (clusters) of statistically correlated water molecules (Bianco and Franzese 2019). According to this mapping, the FS model undergoes a percolation transition along a $T$–$P$ locus that is numerically consistent with the line of extrema of $\xi^{Max}$. A detailed cluster analysis reveals that at high $P$, along the LLPT line, the percolation transition is due to the building up of the HB network. At lower $P$, the origin of the percolation transition is related to the local tetrahedral reordering of the water molecules.

This percolation approach allows us to compute the connectivity length $\xi_C$ of clusters of molecules, which, by definition, diverges along the percolation line at any $P$. Hence, as pointed out by the authors, the rigorous equivalence $\xi \sim \xi_C$ holds only at the critical pressure, but not at other pressures where $\xi$ does not diverge, while $\xi_C$ does (Bianco and Franzese 2019, Coniglio and Klein 1980, Kasteleyn and Fortuin 1969). Nevertheless, it shows that $\xi$ diverges at the LLCP and grows approaching the line of diverging $\xi_C$ for both decreasing and increasing $T$ at low $P$.

### 5.3 The effect of cooperativity in the phase diagram

The FS phase diagram depends on the ratio between the directional and cooperative components of the HB, encoded by the parameters $J$ and $J_\sigma$, respectively. Stokely et al. (2010a) show that tuning $J_\sigma/J$ of the FS model accounts for all the scenarios proposed for the origin of the water anomalies.

By setting $J_\sigma = 0$, i.e., with no cooperativity among the HBs, the FS model reproduces the *SF scenario* for any $J$. For $J_\sigma/J > 0$, but close to zero, the model presents a LLPT ending in a LLCP at $P>0$, as in the *LLCP scenario*. By increasing the ratio $J_\sigma/J > 0$, the LLCP moves from positive pressures and low temperatures to negative pressures and larger temperatures, until it crosses the limit of stability of the liquid phase respect to the gas phase, as in the *CPF scenario*. In this last case, the (positively sloped) liquid-to-gas spinodal merges with the (negatively sloped) liquid-to-liquid spinodal, determining the reentrance of the resulting spinodal for the liquid, as in the *SL conjecture* (Speedy 1982). Stokely et al. (2010a) show that the hypothesis of the LLCP at positive pressure is the only scenario consistent with estimates made from experimental data on the structural and dynamical properties of liquid water (Chumaevskii and Rodnikova 2003, Eisenberg and Kauzmann 1969, Heggie et al. 1996, Henry 2002, Suresh and Naik 2000).

### 7. Summary and future perspectives

In this chapter, we presented the Hamiltonian coarse-grained approach proposed by Franzese and Stanley (FS) in 2002 (Franzese and Stanley 2002a,b) for water under confinement. The FS model coarse-grains the configurations of water molecules and describes the water hydrogen bonding (HB) using two terms, accounting for (i) the directional and (ii) the cooperative components of the HB interaction. A percolation mapping allows the implementation of a cluster Monte Carlo algorithm for an efficient equilibration of a water monolayer at extreme temperatures and pressures, in regions where atomistic water models freeze in a glassy state.

First, we reviewed the dynamical properties of FS water monolayers. The model displays two dynamic crossovers for the HB time-correlation function: the first at $T \sim 252$ K between two non-Arrhenius regimes, and the second at $T \sim 181$ K from non-Arrhenius to Arrhenius behavior (Kumar et al. 2008c, Mazza et al. 2011). These dynamical crossovers, experimentally observed in protein hydration water (Mazza et al. 2011), originate from the maxima of the fluctuations in the HB network, due to the directional and cooperative HB components. Moreover, the FS model reproduces the anomalous diffusion upon pressurization (de los Santos and Franzese 2012).

Second, we reviewed the phase diagram of the FS water (Bianco and Franzese 2014). The model reproduces the increase of density and energy fluctuations upon cooling and the existence of temperatures of maximum and minimum density along isobars. Thermodynamic response functions –the isobaric specific heat, the isothermal compressibility, and the thermal expansivity– display loci of strong and weak maxima. These two sets of maxima correspond to the two dynamic crossovers discussed above. In the supercooled region of the phase diagram, a first-order phase transition separates a high-density liquid (HDL) at higher *P* from a low-density liquid (LDL) at lower *P*. This transition line ends in a critical region (where the loci of thermodynamic maxima converge), where the finite-size scaling analysis of the proper order parameter indicates the presence of a liquid-liquid critical point (LLCP) belonging to the 2D-Ising universality-class. From the LLCP stems the Widom line (WL) calculated as the locus of maxima of the water spatial correlation length $\xi$. The WL coincides with the locus of strong maxima of the response functions in a region that is not accessible to atomistic models. Its geometrical description provides further insight into the building up of the HB network (Bianco and Franzese 2019).

Interestingly, by tuning the ratio between the directional and cooperative components of the HB, the FS model reproduces the proposed scenarios to explain the origin of the water anomalies (Stokely et al. 2010a). When the ratio is zero, no singularity is observed in the thermodynamic quantities at low temperatures, as in the *Singularity Free* scenario. By increasing the ratio, a transition line ending in a critical point appears, as in the LLCP scenario. The larger the ratio, the lower the LLCP pressure (and the larger the LLCP temperature), which passes from positive to negative until the liquid-to-gas stability limit is reached, as in the *Critical Point Free* scenario, with a retracing liquid spinodal, as in the *Stability Limit* conjecture.

These findings, confirmed also by preliminary results for the bulk FS model (Coronas et al. 2016), offer valuable contributions to the ongoing debate on the thermodynamic scenarios for supercooled water and the characterization of the HB network. We believe that the approach of the FS model, by keeping a molecular description of the HB dynamics but strongly reducing the computational cost to sample huge systems, could represent a valuable tool to tackle problems of biological relevance (Bianco et al. 2019a, Bianco and Franzese 2015, Bianco et al. 2019b, 2017a,b) and to explore the design of water-adapted bio-materials (Bianco et al. 2017a, Cardelli et al. 2017, 2019, 2018, Nerattini et al. 2019).

**Acknowledgements**

We thank C. A. Angell, M. Anisimov, M. Bernabei, F. Bruni, C. Calero, D. Chandler, I. Coluzza, F. Coupin, P. Debenedetti, C. Dellago, P. Gallo, P. Kumar, M. Mazza, F. Martelli, P. Poole, S. Sastry, F. Sciortino, F. Starr, K. Stokely, E. Strekalova, A. Zantop, L. Xu for helpful discussions. In particular, we thank H.E. Stanley for introducing us to this field and for his many contribution to it. L. C. acknowledges support by grant n. 5757200 (APIF_18_19-University of Barcelona). O.V. acknowledges support by the Barcelona University Institute of Nanoscience and Nanotechnology (IN2UB). V. B. acknowledges the support from the European Commision through the Marie Sklodowska-Curie Fellowship No. 748170 ProFrost. F. S. acknowledges support by Consejería de Conocimiento, Investigación y Universidad, Junta de Andalucía and European Regional Development Found (ERDF), ref. SOMM17/6105/UGR and Spanish Ministry MINECO project FIS2017-84256-P. G. F. acknowledges support by ICREA Foundation (ICREA Academia prize) and Spanish grant PGC2018-099277-B-C22 (MCIU/AEI/ERDF).